\documentclass[prc,twocolumn,showpacs,amsmath,nofootinbib,superscriptaddress,amssymb]{revtex4-1}
\usepackage{graphicx}
\usepackage{float}
\usepackage{longtable}
\usepackage{array}  
\usepackage{color}
\usepackage{ulem}
\usepackage{isotope}
\usepackage{hyperref}

\newcommand{\Pke}{$P({\bf k},E)$}
\begin{document}

\title{Determination of the proton spectral function of \isotope[12][]{C} from $(e,e^\prime p)$ data}

\author{Artur M. Ankowski}
\affiliation{Institute of Theoretical Physics, University of Wroc\l aw, 50-204 Wroc\l aw, Poland}
\author{Omar Benhar}
\affiliation{INFN, Sezione di Roma , I-00185 Roma, Italy}
\author{Makoto Sakuda}
\affiliation{Physics Department, Okayama University, Okayama 700-8530, Japan}

\date{\today}

\begin{abstract}

The determination of the nuclear spectral function from the measured cross section of  the electron-nucleus scattering process $e + A \to e^\prime + p + (A-1)$
is discussed, and illustrated for the case of a carbon target.
The theoretical model based on the local density approximation, previously employed to derive the spectral function from a combination of accurate 
theoretical calculations and experimental data, has been developed further by including additional information obtained 
from measurements performed with high missing energy resolution. The  implications for the analysis of $\gamma$-ray emission 
associated with nuclear deexcitation are considered.

 \end{abstract}

\maketitle

%%%%%%%%%%%%%%%%%%%%%%%%%%%%%%%%%%%%%%%
\section{Introduction}
\label{introduction}
%%%%%%%%%%%%%%%%%%%%%%%%%%%%%%%%%%%%%%%

The electron-induced proton knockout reaction, in which the scattered electron and the outgoing nucleon are 
detected in coincidence, has long been recognised as a natural and powerful tool for the experimental determination of
the spectral function describing the energy and momentum distribution of protons in the target nucleus; for a short historical account of the developments of this field of research, see Ref.~\cite{NPN}.

Early studies of the $(e,e^\prime p)$ cross sections provided clear cut evidence of the validity of the
mean field approximation underlying the nuclear shell model, exposing at the same time 
its conspicuous limitations~\cite{frullani_mougey}. The spectroscopic lines corresponding to knockout of protons 
occupying the shell model orbitals were, in fact,  clearly observed in the measured missing energy spectra, but 
the corresponding integrated strengths, providing information on the normalisation of the proton wave functions, turned out to be 
significantly lower than expected, regardless of the nuclear target.

In the 1970s, the spectral functions of nuclei ranging from \isotope[12][]{C} to \isotope[58][]{Ni} have been obtained from measurements carried out 
by Mougey {\it et al.} using the 600~MeV electron beam available at CEN Saclay~\cite{saclay}. These data\textemdash spanning the kinematic 
region corresponding to  missing momentum $p_m~<~320$~MeV and missing energy $E_m~<~80$~MeV\textemdash have been later 
combined with  the results of accurate nuclear matter  
calculations~\cite{PkE:nm}, to model the full $p_m$ and $E_m$ dependence of the spectral functions within the framework of the Local Density Approximation (LDA)~\cite{LDA}.

The spectral function \Pke\textemdash trivially related to the two-point Green's function~\cite{Green}\textemdash  is an {\it intrinsic} property of the target nucleus, 
describing the probability of removing a particle of momentum ${\bf k}$ from its ground state leaving the residual system 
with energy $E+E_0-m$, with $E_0$ and $m$ being the ground-state energy and the mass of the knocked out nucleon, respectively. As a consequence, it is a fundamental tool for the analysis of a variety of nuclear scattering processes in the kinematic region in which factorisation of the transition amplitude is expected to be applicable. Notable applications include 
the electron- and neutrino-nucleus scattering cross sections; see, e.g., Refs.~\cite{LDA,Benhar:2006wy,Benhar:2005dj,Benhar:2006nr,Ankowski:2014yfa,gamma:16O}. 
It should be pointed out, however, that different reaction mechanisms and different kinematic setups are sensitive to different features of the spectral function.
Theoretical analyses of inclusive $(e,e^\prime)$ cross sections, involving integration over the phase space of the knocked out nucleon,  
turn out to be largely unaffected by  the details of the energy dependence  of \Pke.  On the other hand, accurate calculations of processes involving transitions
between energy levels predicted by the nuclear shell model require a detailed description of the removal energy distributions, allowing to pin down the relevant spectroscopic strengths.
 
The authors of Ref.~\cite{gamma:16O} applied the spectral function formalism in a study of $\gamma$-ray emission from deexcitation
of the residual nucleus produced in \isotope[16][8]{O}$(\nu,\nu^\prime N)$ reactions,  with $N$ denoting either a proton or a neutron,  which 
provides a signature of neutral-current neutrino-nucleus interactions in water-Cherenkov detectors. Their analysis employed the oxygen spectral function  
of  Benhar {\it et al}.~\cite{Benhar:2005dj}, obtained from the LDA-based approach of Ref.~\cite{LDA} using the Saclay data  
of Bernheim {\it et al.}~\cite{Bernheim:1981si}. This spectral function model allows the determination of the spectroscopic strengths 
of the oxygen $p$-states, the values of which turn out to be in remarkable agreement with those obtained from the analysis of the high-resolution missing energy 
spectra of the \isotope[16][8]{O}$(e,e^\prime, p)$ process performed by Leuschner {\it et al.}~\cite{nikhef:oxygen}.

In this article, we discuss the derivation of an improved model of the carbon spectral function of Ref.~\cite{LDA}, designed to describe the 
missing energy dependence of the  \isotope[12][6]{C}$(e,e^\prime, p)$ cross section measured at NIKHEF using a 500 MeV high-duty-cycle 
electron beam and a pair of high-resolution spectrometers~\cite{nikhef}. The improved spectral function model makes it possible to pin down the spectroscopic 
strengths of the shell model states of the carbon ground state, which are needed to extend the analysis of Ref.~\cite{gamma:16O} to the case of neutral-current neutrino interactions in liquid scintillator detectors. 

The reminder of this article is organised as follows. The decomposition into pole and continuum contributions
and the phenomenological procedure employed to obtain the nuclear spectral function within LDA are discussed in Sections~\ref{SF:analytic} and \ref{SF:eep}, respectively, whereas Section~\ref{SF:12C} describes the derivation of an improved model of the carbon spectral function, including the information 
provided by the high-resolution missing energy spectrum reported in Ref.~\cite{nikhef}. Finally, in Section~\ref{summary} we summarise the results of our work  and outline their applications.   

%%%%%%%%%%%%%%%%%%%%%%%%%%%%%%%%%%%%%%%
\section{The Nuclear Spectral function}
\label{SF}
%%%%%%%%%%%%%%%%%%%%%%%%%%%%%%%%%%%%%%%

When a nucleon is knocked out from a nucleus,  the residual $(A-1)$-particle system
may be  left in either a bound or a continuum state, in which at least one of  the spectators is excited to a positive energy  level. 
These two reactions mechanisms give rise to distinctively different missing energy distributions. 

%%%%%%%%%%%%%%%%%%%%%%%%%%%%%%%%%%%%%%%
\subsection{Analytic properties of the missing energy distribution}
\label{SF:analytic}
%%%%%%%%%%%%%%%%%%%%%%%%%%%%%%%%%%%%%%%

In $(e,e^\prime p)$ processes,  the missing energy is defined as
\begin{align}
\label{missing:energy}
E_m = \omega - T_p - T_{A-1} \approx \omega - T_p \ , 
\end{align}
where $\omega=E_e-E_{e^\prime}$ denotes the electron energy loss, while  $T_p$ and $T_{A-1}$, with typically $T_{A-1} \ll T_p$,  
are the kinetic energies of 
the outgoing proton and the residual nucleus, respectively. Equation~\eqref{missing:energy} shows that, to the extent to which 
Final State Interactions (FSI) between the struck nucleon and the recoiling system  can be neglected or reliably described by an optical potential, the {\it measured} value of $E_m$  can be identified 
with the removal energy of the knocked out particle. It follows that transitions to bound ($A-1$)-nucleon states\textemdash associated with 
one-particle\textendash one-hole excitations of the target nucleus\textemdash lead to the appearance of 
a missing energy distribution consisting of narrow peaks corresponding to the poles of the spectral function at 
$E_m \approx |E_\alpha|$, with $E_\alpha<0$ being the  
energy of the shell model state $\alpha$. On the other hand, a smooth distribution, extending into the region of missing energies above 
the two-nucleon emission threshold, signals the occurrence of $n$-particle\textendash $n$-hole excitations, with $n\geq 2$, arising from strong correlations among  the nucleons in the target ground state. 

Based on the above considerations, the proton spectral function can be written in {\it model-independent} form as  
\begin{align}
\label{KL:decomposition}
P({\bf k}, E) = P_{\rm MF}({\bf k}, E)  + P_{\rm corr}({\bf k}, E)  \  , 
\end{align}
where $P_{\rm MF}$ and $P_{\rm corr}$ denote the pole, or mean-field (MF), and continuum,  or 
correlation (corr), contributions, respectively. Integration over the full energy and momentum ranges yields the normalisation
\begin{align}
\label{normalisation}
\int d^3k dE~P({\bf k},E) = Z \ ,  
\end{align}
with $Z$ being the nuclear charge number.

According to the shell model, in which correlations are disregarded altogether, $P_{\rm corr}({\bf k}, E) = 0$, and the 
spectral function reduces to 
\begin{align}
\label{shell:model}
 P_{\rm SM}({\bf k}, E)  = \sum_{\alpha \in \{ F \}} |\phi_\alpha({\bf k})|^2 \delta(E-|E_\alpha|)  \  .
\end{align}
In the above equation, $\alpha \equiv \{ n_\alpha,\ell_\alpha,j_\alpha \}$ denotes the set of quantum numbers 
specifying the proton state $\alpha$, $\phi_\alpha({\bf k})$ is the corresponding unit-normalised wave function in momentum space, and 
 the sum is extended to all states belonging to the Fermi sea $\{ F \}$. 

The right-hand side of Eq.~\eqref{shell:model} is substantially modified by the effects of nuclear dynamics beyond the shell model. Ground-state correlations
lead to the appearance of a non-vanishing continuum contribution,  which turns out to account for $\sim$20\% of the normalisation~\cite{PkE:nm}.  
Furthermore, the pole contribution\textemdash the normalisation of which is obviously reduced by the same amount with respect to the shell-model prediction\textemdash is modified by the coupling between one-hole and two-hole\textendash one-particle states of the recoiling nucleus, giving rise to a broadening of the energy distribution of the shell model states~\cite{Z}. The resulting expression can be conveniently written in the form
\begin{align}
P_{\rm MF}({\bf k}, E)  & = \sum_{\alpha \in \{ F \}} | \phi_\alpha({\bf k})|^2 f_\alpha(E)  \ , 
\label{PkE:MF}
\end{align}
with the normalisation of the finite-width distribution $f_\alpha(E)$ being given by 
\begin{align}
\label{def:Z}
\int dE \ f_\alpha(E) = Z_\alpha \ .
\end{align} 

The decomposition of the spectral function into discrete and continuum contributions, which follows immediately from the K\"allen-L\'ehman representation of the 
two-point Green's function~\cite{FW}, is a powerful method for the model-independent identification of correlation effects, the size of which 
is gauged by the difference $[1 - ( \sum_{\alpha}  Z_\alpha)/Z] \geq 0$.  A detailed discussion of this topic can be found in Ref.~\cite{Z}.

It should be kept in mind that, although Eqs.~\eqref{KL:decomposition} and \eqref{PkE:MF} provide a useful and physically motivated scheme for the parametrisation  
of the full spectral function, the width of the distributions $f_\alpha(E)$, which is known to be  vanishingly small for valence states in the vicinity of 
the Fermi surface, becomes very large for $E_\alpha \ll E_F$, with $E_F$ being the proton Fermi energy. As a consequence, deeply bound states 
turn out to be very short-lived, and their interpretation in terms of shell model orbits must be carefully analysed.

Theoretical calculations of the nuclear spectral function based on advanced microscopic models of nuclear dynamics and accurate 
computational techniques have been carried out for the three-nucleon system~\cite{Dieperink:1976wy,CiofidegliAtti:1980dbw,Meier-Hajduk:1983pii}, 
the doubly closed shell nucleus \isotope[16][8]{O}~\cite{Geurts:1996zza,Rijsdijk:1996zz,Polls:1996hg,noemi:1}, and isospin-symmetric nuclear matter~\cite{PkE:nm,Ramos:1989hqs}.
Calculations involving a somewhat simplified treatment of the continuum component have been performed for \isotope[4][]{He}~\cite{CiofidegliAtti:1990dh,Morita:1991ka,Benhar:1993ja}.
More recently, the spectral functions of nuclei as heavy as \isotope[40][20]{Ca} or having  significant neutron excess, such as \isotope[40][18]{Ar}, have been also studied using the Green's function Monte Carlo method~\cite{noemi:2}. 

%%%%%%%%%%%%%%%%%%%%%%%%%%%%%%%%%%%%%%
\subsection{Phenmomenological determination of the spectral function}
\label{SF:eep}
%%%%%%%%%%%%%%%%%%%%%%%%%%%%%%%%%%%%%%

In the absence of FSI, the cross section of the process $e + A \to e^\prime + p + (A-1)$,  can be written in the simple and transparent 
form
\begin{align}
\label{eep:xs}
\frac{d\sigma_A}{dE_{e^\prime} d\Omega_{e^\prime} dE_{p^\prime} d\Omega_{p^\prime} }  = E_{{p^\prime}} |{\bf p}^\prime| \   \sigma_{ep} \ P(-{\bf  p}_m, E_m) \ . 
\end{align}
Here $\Omega_{e^\prime}$ and $\Omega_{p^\prime}$ are the solid angles specifying the directions of 
the scattered electron and the outgoing proton, whose energy and momentum are denoted $E_{{p^\prime}}$ and ${\bf p}^\prime$, respectively. The 
cross section $\sigma_{ep}$ describes elastic electron scattering off a bound proton of momentum $-{\bf p}_m$ and removal energy $E_m$. The missing momentum is defined as ${\bf p}_m = {\bf q} - {\bf p}^\prime$, with ${\bf q}$  being the momentum transfer.

The large data set of precisely measured $(e,e^\prime p)$ cross sections collected at Saclay and NIKHEF from the 1970s to the  1980s
covers  mainly  the kinematic region where mean-field dynamics dominates. As a consequence, Eq.~\eqref{eep:xs} implies that, as long as the effects of FSI can be reliably taken into account, these data have the potential to 
allow an accurate determination of the spectroscopic strengths $S_\alpha$, obtained from integration of the measured missing energy spectra over the regions $|E_\alpha| -\Gamma_\alpha \lesssim E_m \lesssim |E_\alpha| + \Gamma_\alpha$,  with $\Gamma_\alpha>0$ being the width of the peak of $f_\alpha(E)$ 
at $E \approx |E_\alpha|$. Note that, by definition, $S_\alpha \leq Z_\alpha$, because the integration of Eq.~\eqref{def:Z} is extended to all values of $E$. 
Therefore, $Z_\alpha$ includes contributions associated with the occurrence of two-hole\textendash one-particle intermediate states of the residual nucleus, giving rise to the appearance of appreciable tails in the distribution $f_\alpha(E)$~\cite{Z}.  

In experimental analyses, the corrections originating from FSI  are usually treated within the well established framework known as Distorted Wave Impulse Approximation (DWIA), based on the 
use of complex proton-nucleus optical potentials; for a detailed discussion of the formalism of DWIA, see, e.g., Refs.~\cite{DWIA:1,DWIA:2}.

Experimental studies aimed at pinning down the correlation contribution  $P_{\rm corr}({\bf k}, E)$ require a careful choice of the kimematic setup, needed 
to minimise the background due to the occurrence of inelastic processes, as well as a more refined treatment of FSI~\cite{daniela_FSI}. The data collected at 
Jefferson Lab using a carbon target, while not providing a detailed description of the full energy and momentum dependence, show a remarkable consistency 
between the measured correlation strength and the missing strength in the mean-field sector observed by previous experiments~\cite{daniela}. 

The formalism based on LDA, developed  by the authors of Ref.~\cite{LDA} in the 1990s, allows to determine the nuclear spectral function by exploiting the decomposition of Eq.~\eqref{KL:decomposition} and the availability of accurate measurements of the mean-field contribution, Eq.~\eqref{PkE:MF}. The main tenet underlying this approach\textemdash strongly supported by the results of both theoretical and 
experimental studies~\cite{Alvioli:2016wwp}\textemdash is that short-range nuclear dynamics, driving the appearance of correlations among the nucleons, are largely unaffected by surface and shell effects. As a consequence, correlation effects can be reliably evaluated in uniform nuclear matter. 

Within LDA, the continuum component of the spectral function of a nucleus of mass number $A$ is approximated by
\begin{align}
\label{LDA:definition}
P_{\rm corr}^{\rm LDA}({\bf k},E) = \int d^3 r \ \varrho_{\rm A}({\bf r})~P_{\rm corr}^{\rm NM}[{\bf k},E; \varrho = \varrho_A({\bf r})] \ , 
\end{align}
where $\varrho_A({\bf r}) $ is the unit-normalised nuclear density distribution, and $P_{\rm corr}^{\rm NM}({\bf k},E; \varrho)$ denotes the continuum component of the spectral function of nuclear matter at density $\varrho$. A detailed derivation of Eq.~\eqref{LDA:definition} can be found in Ref.~\cite{LDA}. 

Accurate calculations of $P_{\rm corr}^{\rm NM}$ in the density range $0.25 \leq (\varrho/\varrho_0) \leq 1$, with $\varrho_0 \approx 0.16 \ {\rm fm}^{-3}$ being the equilibrium density of isospin-symmetric matter, have been carried out using an advanced 
many-body approach and a realistic model of the nuclear Hamiltonian~\cite{PkE:nm,LDA}. 

%%%%%%%%%%%%%%%%%%%%%%%%%%%%%%%%%%%%%
\section{The proton spectral function of carbon}
\label{SF:12C}
%%%%%%%%%%%%%%%%%%%%%%%%%%%%%%%%%%%%%

%%%%%%%%%%%%%%%%%%%%%%%%%%%%%%%%%%%%%
\subsection{LDA model based on Saclay data}
\label{12C:saclay}
%%%%%%%%%%%%%%%%%%%%%%%%%%%%%%%%%%%%%
Mougey {\it et al.} obtained the mean-field contribution to the proton spectral function of carbon from the $(e,e^\prime p)$ data reported in Ref.~\cite{saclay}. 
Based on the combined 
analysis of the missing momentum and missing energy dependence of the measured cross section they were able to single out  the contribution of proton knockout 
from different shell-model states. The peak observed at $15 \leq E_m \leq 21.5$ MeV, associated with a maximum of the corresponding momentum 
distribution at $p_m \approx 100$ MeV, was attributed to the valence $p$-state, whereas the strength at $30 \leq E_m \leq 50$ MeV, whose $p_m$-dependence features a maximum at $p_m= 0$, was interpreted as an $s$-state, that is, $\ell=0$, contribution. This decomposition allowed to extract the spectroscopic strengths of the $p$- and $s$-states from the data, corrected to take into account the effects of FSI. 

\begin{figure}[th]
\includegraphics[width=0.95\columnwidth]{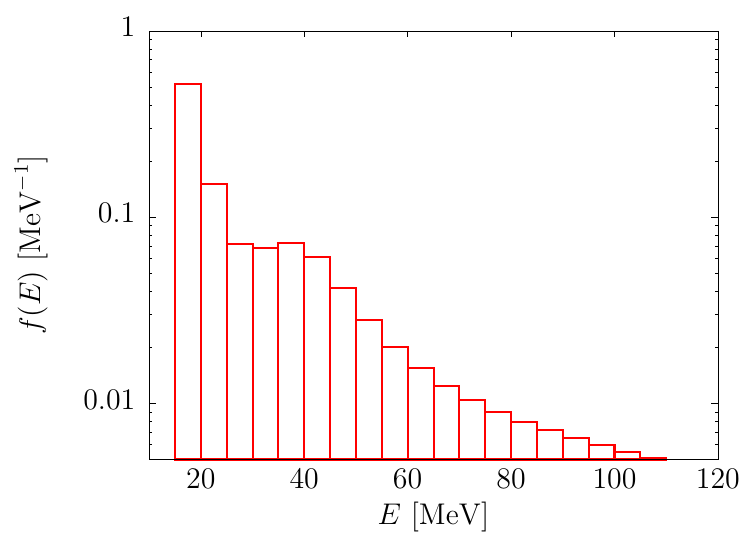}
\vspace*{-.10in}
\caption{\label{FE:LDA} Energy distribution obtained from momentum integration of the LDA spectral function of Ref.~\cite{LDA}.}
\end{figure}

The authors of Ref.~\cite{LDA} have combined the empirical information reported in Ref.~\cite{saclay} with the nuclear matter results of Ref.~\cite{PkE:nm} to 
obtain the proton spectral function of carbon within LDA.  The corresponding energy distribution, defined as
\begin{align}
\label{edist:LDA}
f_{\rm LDA}(E) = \int d^3k~ \big[  P_{\rm MF}({\bf k},E)  + P_{\rm corr}^{\rm LDA}({\bf k},E) \big] \ , 
\end{align} 
with $P_{\rm MF}$ and $P_{\rm corr}^{\rm LDA}$ given by Eqs.~\eqref{PkE:MF} and \eqref{LDA:definition}, respectively, is displayed in Fig.~\ref{FE:LDA}.
Note that $f(E)$ is nonzero in the region  $E \geq E_{\rm thr}$, with the proton knockout threshold $E_{\rm thr}$ being defined as
\begin{align}
E_{\rm thr} =  | M( \isotope[12][]{C}) - M(\isotope[11][]{B}) - m_p - m_e | \ , 
\label{def:Ethr}
\end{align}  
where $m_p=938.27$~MeV and $m_e=0.51$~MeV denote the proton and electron mass, respectively~\cite{PDG}. 
Using $M( \isotope[12][]{C})~=~11177.93$~MeV and $M( \isotope[11][]{B})~=~10255.11$~MeV~\cite{NDT}, Eq.~\eqref{def:Ethr} yields 
$E_{\rm thr} = 15.96$ MeV.

Integration of the distribution of Fig.~\ref{FE:LDA} yields the spectroscopic factors $S_p = 2.8$ and $S_s = 1.2$, corresponding to occupation 
probabilities $S_\alpha/(2j_\alpha + 1)$ = 0.7 and 0.6,  respectively. % to be compared with the values  $S_p= 2.5$ and $S_s=1$ reported in Ref.~\cite{saclay}.  
Note that the shell-model predicts $S_p= 4$ and $S_s=2$.

%%%%%%%%%%%%%%%%%%%%%%%
%%%%%%%%%%%%%%%%%%%%%%%%

%%%%%%%%%%%%%%%%%%%%%%%%%%%%%%%%%%%%%%%
\subsection{Parametrisation of NIKHEF data}
\label{12C:NIKHEF}
\begin{figure}[ht]
\includegraphics[width=0.95\columnwidth]{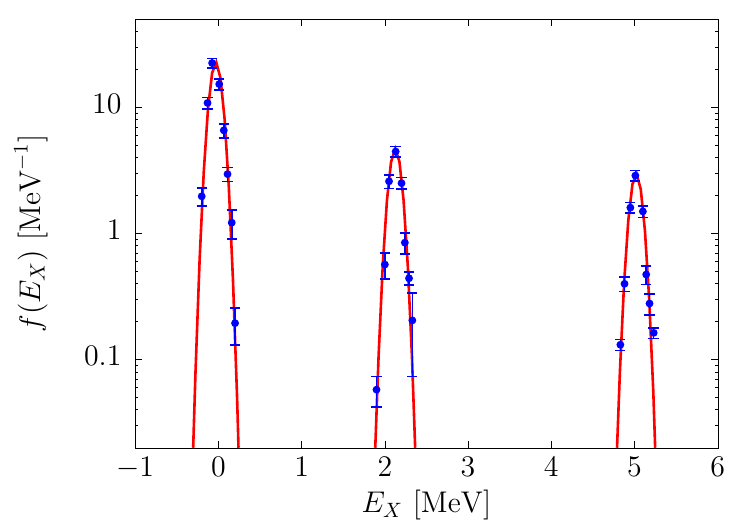}
\vspace*{-.10in}
\caption{\label{fig1} Comparison between the energy distributions extacted from NIKHEF data~\cite{nikhef}
and the results of the gaussian fit employed in this work, represented by the solid lines.}
\end{figure}

%%%%%%%%%%%%%%%%%%%%%%%%%%%%%%%%%%%%%%%
In Ref.~\cite{nikhef}, Van der Steenhoven {\it et al.} reported the missing energy spectrum of the  \isotope[12][]{C}$(e,e^\prime p)$ reaction in the range corresponding to knock out of a proton occupying the valence $p$-states of  carbon, measured at NIKHEF with 150 keV resolution (FWHM). 
The resulting distribution, $f(E_X)$, displayed in Fig.~\ref{fig1}, exhibits three prominent peaks  at excitation energy $E_X = E_m - E_{\rm thr} =$ 0\textemdash that is, the process 
in which the recoiling nucleus of  \isotope[11][]{B}  is left in its ground state\textemdash a $p_{1/2}$ state at $E_X = 2.125$ MeV, 
and a $p_{3/2}$ state at  $E_X = 5.020$~MeV. 
The primary aim of this work is improving the spectral function model or Ref.~\cite{LDA}, based on Saclay data, by inclusion of the additional 
information provided by the data of Ref.~\cite{nikhef}.

As a first step,  the distribution of Fig.~\ref{fig1} has been modeled assuming gaussian-shaped peaks whose position and width, as well as the 
corresponding spectroscopic strengths, have been determined using a code based on the Marquardt fitting algorithm~\cite{ingo_code}.

%%%%%%%%%%%%%%%%%%%%%%
%%%%%%%%%%%%%%%%%%%%%%%%

In order to combine NIKHEF data with the LDA spectral function based on Saclay data, 
the fit shown by the solid line of Fig.~\ref{fig1}, hereafter 
$f_{\rm fit}$, has been renormalised in such a way as to match the $p$-state normalisation resulting from integration in the LDA spectral function 
of Ref.~\cite{LDA} in the range $15 \leq E_{\rm m} \leq 21.5$ MeV, which will be denoted $\Delta E$. 
After renormalisation, the fitted energy distribution fulfils the condition
\begin{align}
 \int_{\Delta E} dE_{\rm m} f_{\rm fit} (E_{\rm m})   = \int_{\Delta E} dE_{\rm m} f_{\rm LDA} (E_{\rm m}) = S_p \ , 
\end{align}
with $S_p = 2.8$, by construction.

In Table~\ref{Z:fit}, the spectroscopic strengths defined as 
\begin{align}
\label{def:S_alpha}
S_{\alpha} = \int_{\Delta{E_\alpha}} dE_{\rm m} f_{\rm fit}(E_{\rm m}) \ ,
\end{align}
as well as the values of the occupation probabilities of the shell-model orbits, $Z_\alpha = S_\alpha/(2j_\alpha +1)$, are compared 
to the experimental results reported in Ref.~\cite{nikhef}. 
Here the index $\alpha$ runs over the states at $E_m =$ 15.96, 18.08, and 20.97 MeV, and $\Delta{E_\alpha}$ is a narrow integration 
region designed to pin down the associated pole strength. 
%%%%%%%%%%%%%%%%
\begin{table}[hbt]
    \begin{tabular}{|l|l|l|l|}
    \hline\hline
    $\alpha$  &  \ \ \ $p_{3/2}$ \ \ \ \ \  &  \ \ \ $p_{1/2}$ \ \ \ \ \ & \ \ \ $p_{3/2}$ \ \ \ \ \ \\[3pt]
    \hline
    $S_\alpha^{\rm expt}$  & 1.72 & 0.26 & 0.20 \\
    \hline
    $Z_\alpha^{\rm expt}$\ \ \ \ \  & 0.43 & 0.13 & 0.05 \\
    \hline
    $S_\alpha$  & 1.747 & 0.273 & 0.201 \\
    \hline
    $Z_\alpha$ \ \ \ \ \   & 0.437 & 0.137 & 0.05 \\
    \hline 
    $E_m$ [MeV] & 15.96 & 18.08 & 20.97 \\
    \hline \hline
    \end{tabular}
    \caption{\label{Z:fit} Spectroscopic strengths and occupation probabilities of the valence $p$-states of \isotope[12][]{C} at missing energy $E_m$, obtained 
    from Eq.~\eqref{def:S_alpha}. The experimental results reported in Ref.~\cite{nikhef} are also listed, for comparison.} 
\end{table}
%%%%%%%%%%%%%%%%

The total strength, $\sum_{\alpha} S_\alpha = 2.22$, should be compared to value, $S_p = 2.8$ obtained from the LDA spectral function of Ref.~\cite{LDA}.
This $\sim$25\% discrepancy is to be ascribed to additional contributions arising from fragmentation of the shell-model states, clearly visible in 
the missing energy spectra reported in Ref.~\cite{nikhef} and not included in the determination of the $S_\alpha$ from Eq.~\eqref{def:S_alpha}.

%%%%%%%%%%%%%%%%%%%%%%%%%%%%%%%%%%%%%%%
\subsection{Improved spectral function model}
\label{12C:combined}
%%%%%%%%%%%%%%%%%%%%%%%%%%%%%%%%%%%%%%%
 
\begin{figure}[h!]
\vspace*{.075in}
\includegraphics[width=0.95\columnwidth]{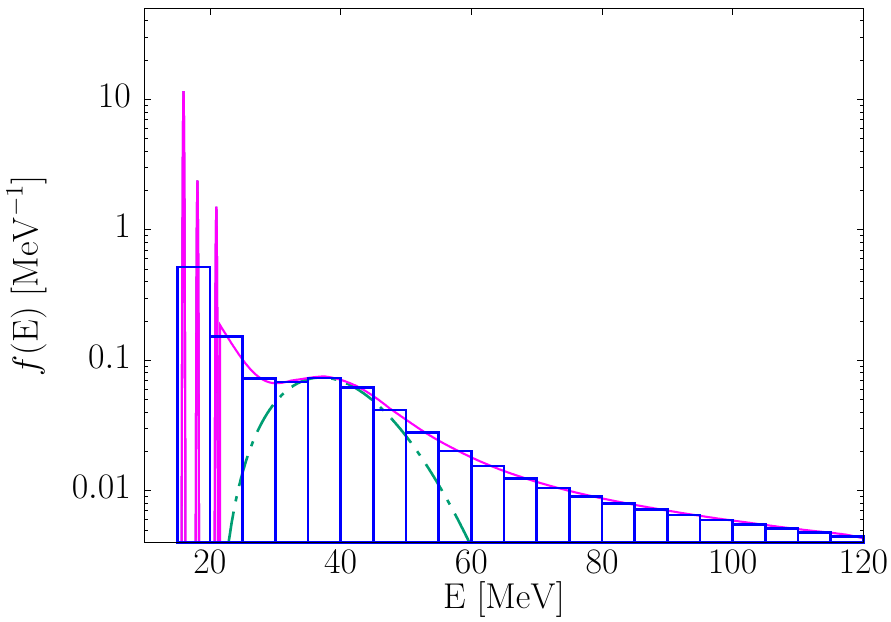}
\vspace*{-.10in}
\caption{The full line represents the energy distribution obtained combining the NIKHEF data at $E \leq E_{\rm match} = 21.5$ MeV and 
the results of Ref.~\cite{LDA} at at $E > E_{\rm match}$, while the dot-dash line corresponds to the $s$-shell contribution described in the text. 
For comparison, the full distribution of Ref.~\cite{LDA} is shown by the histogram.\label{edist:combined} }
\end{figure}

The energy distribution assembled combining the fit to NIKHEF data, spanning the region $E~\leq~E_{\rm match} = 21.5$ MeV, 
and the LDA distribution of Ref.~\cite{LDA} at $E > E_{\rm match}$ is displayed by the thick solid line of Fig.~\ref{edist:combined}. 
Comparison with the histogram representing the full distribution of Ref.~\cite{LDA}, corresponding to Eq.~\eqref{edist:LDA}, highlights the importance of 
taking into acount the results of measurements performed with high missing energy resolution.  

The integrated $s$-shell strength at $E \geq 21.5$ MeV discussed by Mougey {\it et al.}~\cite{saclay} has been also modeled, assuming a Maxwell-Boltzmann shape 
peaked at $E = 37$ MeV. This component, shown by the dot-dash line of Fig.~\ref{edist:combined}, contributes a spectroscopic strength  
$S_s = 1.15$, which turns out to be within $\sim 5$\% of the corresponding result obtained from the data of Ref.~\cite{saclay}. 

%%%%%%%%%%%%%%%%%%%%%%
%%%%%%%%%%%%%%%%%%%%%%%
  
A procedure analogue to that used for deriving the energy distribution has been employed to obtain an improved model 
of the full spectral function.
Within this scheme, $P({\bf k},E>E_{\rm match})$ is the same as the LDA spectral function of Ref.~\cite{LDA}, while in the region $E~\leq~E_{\rm match}$
we assume the factorised form
\begin{align}
\label{pkelow}
P({\bf k},E<E_{\rm match}) = n_p({\bf k}) f(E)  \ , 
\end{align}
where $n_p({\bf k}) = | \phi_p({\bf k}) |^2$\textemdash with $\phi_p({\bf k})$ being the $p$-state wave function in momentum space\textemdash  and $f(E)$ is the distribution represented by the solid line of Fig.~\ref{edist:combined}. 
The energy and momentum dependence of the resulting spectral function, normalised according to Eq.~\eqref{normalisation}, is illustrated in Fig.~\ref{SF}.

%%%%%%%%%%%%%%%%%%%%%%
\begin{figure}[h]
\vspace*{.075in}
\includegraphics[width=0.95\columnwidth]{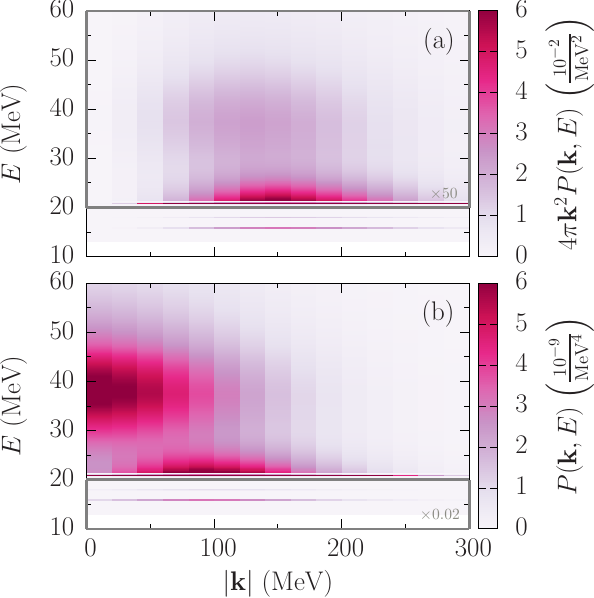}
\vspace*{-.10in}
\caption{Energy and momentum dependence of the carbon spectral function 
derived from the approach described in this work. 
Panels (a) and (b) have been obtained with and without including 
the geometric factor $4\pi{\bf k}^2$, respectively.\label{SF}}
\end{figure}
%%%%%%%%%%%%%%%%%%%%%%%%

\begin{figure}[h]
\vspace*{.075in}
\includegraphics[width=0.95\columnwidth]{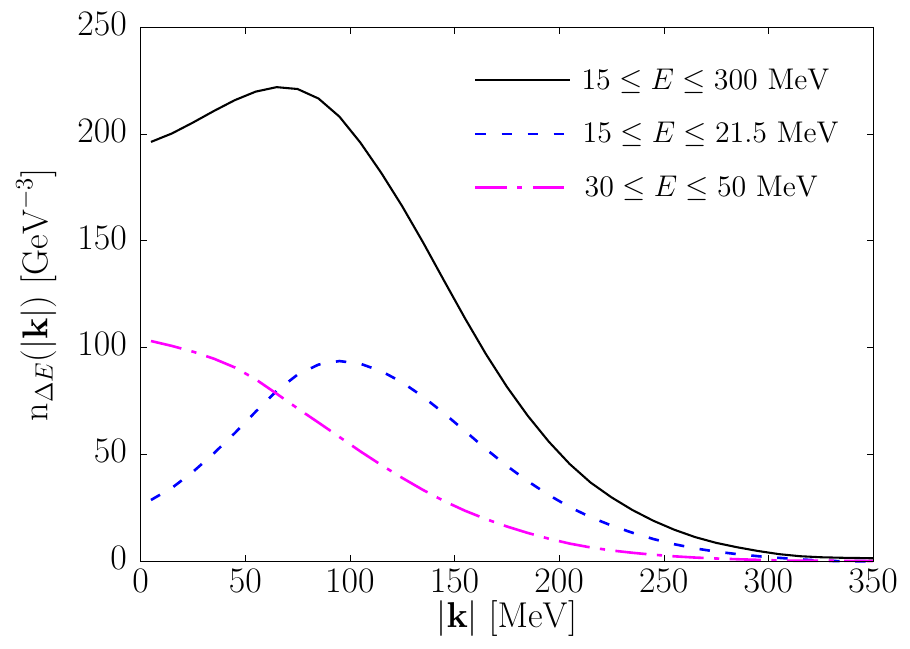}
\vspace*{-.10in}
\caption{Partial proton momentum distributions of carbon, obtained from Eq.~\eqref{define:momdis} using the spectral function derived in this work and 
different integration regions $\Delta E$. The spectroscopic factors resulting from integration of the dashed and dot-dash lines, corresponding to $p$- and  
$s$-shell strength, respsctively, turn out to be $S_p = 2.8$ and $S_s = 1.2$.
\label{momdis1} }
\end{figure}

Integration of the spectral function over the energy range $\Delta E$ yields the partial momentum distribution 
\begin{align}
\label{define:momdis}
n_{\Delta E}({|\bf k|}) = \int_{\Delta E} dE~P({\bf k},E) \ .   
\end{align} 
The above definition implies that suitable choices of $\Delta E$ allow to pin down the contributions of protons occupying 
different shell-model states, or belonging to a correlated pair in the continuum. Figure~\ref{momdis1} shows 
that the partial distributions corresponding to $30 < E < 50$ MeV and $15 < E < 21.5$ MeV peak at $|{\bf k}| = 0$ and $\sim$100 MeV, 
as expected for $s$- and $p$-shell protons, respectively, and  become vanishingly small at momenta larger than the
typical nuclear Fermi momentum $k_F \approx250$ MeV. On the other hand, integration over the full energy range, extending 
up to $E=300$ MeV, leads to the appearance of a high momentum tail originating from strong  short-range correlations among the nucleons. 

%%%%%%%%%%%%%%%%%%%%%%
%%%%%%%%%%%%%%%%%%%%%%%
 
\begin{figure}[h]
\vspace*{.075in}
\includegraphics[width=0.95\columnwidth]{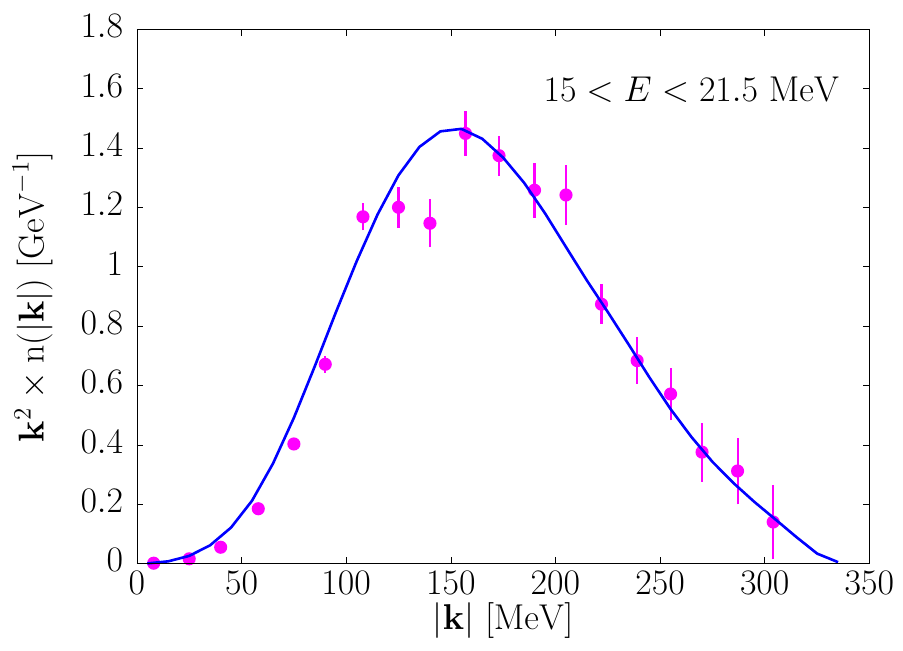}
\vspace*{-.10in}
\caption{\label{momprob1} Momentum probability of the $p$-state of \isotope[12][]{C}, obtained 
from integration of the spectral function described in this work in the range $15 < E < 21.5$ MeV.
The effects of FSI have been taken into account using the same correction applied to the data by the authors of Ref.~\cite{saclay}.}
\end{figure}

In Fig.~\ref{momprob1}, the $p$-state momentum probability obtained multiplying by a factor ${\bf k}^2$ the corresponding 
distribution\textemdash shown by the dashed line of Fig.~\ref{momdis1}\textemdash is compared to the experimental result reported in 
Ref.~\cite{saclay}. Note that, in order to perform a meaningful comparison, the distribution obtained from our model has been modified to 
account for the effects of FSI, applying the same correction employed in the analysis of Mougey {\it et al.}~\cite{saclay}. The agreement between 
experimental data and the results of our approach turns out to be remarkable.

%%%%%%%%%%%%%%%%%%%%%%
%%%%%%%%%%%%%%%%%%%%%%%%

%%%%%%%%%%%%%%%%%%%%%%%%%%%%%%%%%%%%%%%
\section{Summary and Outlook}
\label{summary}
%%%%%%%%%%%%%%%%%%%%%%%%%%%%%%%%%%%%%%%

We have updated the carbon spectral function of Ref.~\cite{LDA}\textemdash obtained from a semi-phenomenological 
model based on LDA\textemdash by including additional information provided by the high-resolution \isotope[12][]{C}$(e,e^\prime p)$ 
experiment performed at NIKHEF, Amsterdam, by Van der Steenhoven {\it et al.}~\cite{nikhef}.

The new spectral function features a modified energy dependence in the range at $E \leq 21.5$ MeV, relevant to removal of a $p$-shell  
proton, and reduces to the $P({\bf k},E)$ of Ref.~\cite{LDA} at larger
The primary improvement lies in the capability to clearly identify the three previously unresolved states with removal energies 15.96, 18.08, and 20.97 MeV, which 
turn out to account for $\sim$55\% of the $p$-state strength predicted by the nuclear shell model. 

The strength in the region $E>21.5$ MeV is described in terms of an $s$-shell contribution\textemdash extending at energies $21.5 \lesssim E 
\lesssim 60$ MeV and 
modeled by a Maxwell-Boltzmann distribution peaked at $E \sim 37$ MeV\textemdash and a smooth background  contribution, arising from fragmentation of 
shell-model states and nucleon-nucleon correlations, extending to energies as high as $\sim 300$ MeV.

It should be pointed out that the modified energy dependence of the new spectral function does not affect significantly 
the inclusive electron scattering cross section, the calculation of which involves integrations of $P({\bf k},E)$ over  
broad energy ranges.
 
%%%%%%%%%%%%%%%%%%%%%%
\begin{figure}[h]
\vspace*{.075in}
\includegraphics[width=0.95\columnwidth]{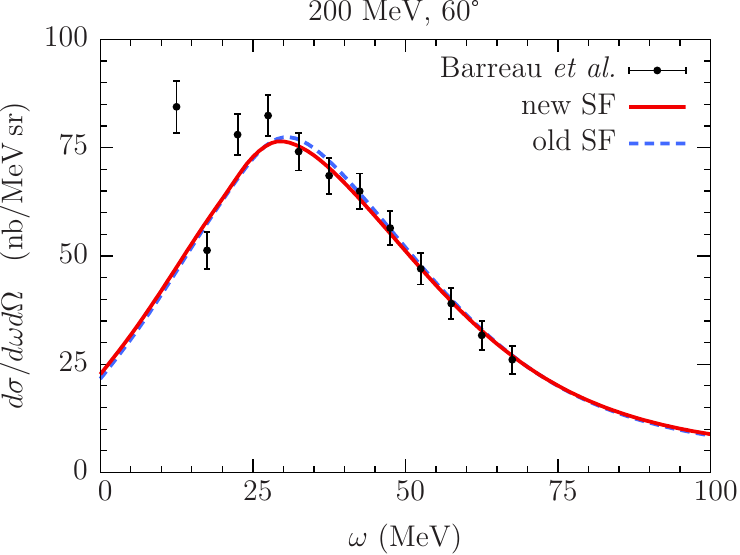}
\vspace*{-.10in}
\caption{Differential \isotope[12][]{C}$(e,e^\prime)$ cross section at beam energy $E_e = 200$ MeV and electron scattering angle 
$\theta_e = 60^\circ$, as a function of electron energy loss $\omega$. Experimental data from Ref.~\cite{Barreau:1983ht}.\label{inclusive}}
\end{figure}
%%%%%%%%%%%%%%%%%%%%%%%%

To illustrate this point at quantitative level, we have compared the inclusive electron-carbon cross section corresponding to beam energy $E_e = 200$ MeV and
electron scattering angle $\theta_e = 60^\circ$ obtained from the spectral function derived in this work to the one computed by the  
authors of Ref.~\cite{Ankowski:2014yfa} using the spectral function of Ref.~\cite{LDA}. It appears that the results of the two calculations 
are nearly indistinguishable from one another over the whole range of electron energy loss $\omega$. 

Note that the failure to reproduce the data at low $\omega$
signals the breakdown of the factorisation approximation underlying the spectral function formalism. In particular, the excitation of the giant dipole resonance at 
$E_X = 22.6$ MeV~\cite{ref:nuclearDataSheets_12}\textemdash corresponding to $\omega = 24.2$ MeV due to nuclear recoil\textemdash is not accounted 
for in the calculations presented in Fig.~\ref{inclusive}. In addition, the experimental data include contributions originating from elastic scattering and transitions
to discrete levels, leading to the appearance of the jump of the cross section at the lowest value of $\omega$. 
 
The potential of the improved $P({\bf k},E)$ can be fully exploited in studies requiring a detailed description of the $E$-dependence, 
such as the analyses of nuclear deexcitation processes. Notable example are the calculations of the rate of $\gamma$-rays emitted by the residual nucleus 
in the aftermath of neutral current neutrino interactions in the quasielastic channel~\cite{gamma:16O}, or the identification of the leading proton kinematics in Monte Carlo cascade model simulations of FSI in neutrino-nucleus interactions~\cite{Ershova:2023dbv}.  The results reported in this article will allow to extend the study of Ref.~\cite{gamma:16O}, relevant to measurements performed using water-Cherenkov detectors, such as K2K~\cite{Kameda:2006zn}, to the analysis of experiments using liquid scintillator detectors,  such as KamLAND~\cite{KamLAND:2022ptk} and JUNO~\cite{JUNO:2015zny}. The need of accurate models of the carbon and oxygen  spectral functions for the description of deexcitation processes in Monte Carlo simulations has been recently discussed in Ref.~\cite{PhysRevD.109.036009}.

\acknowledgements
A.M.A. and O.B. gratefully acknowledge partial support by grant NSF PHY-1748958 to the Kavli Institute for Theoretical Physics (KITP), where 
this research was initiated. The work of A.M.A. is partly supported by the National Science Centre under grant UMO-2021/41/B/ST2/02778. 
\bibliographystyle{apsrev4-1}

\end{document}